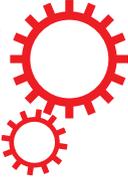

# OPEN

# Gut Dysbiosis and Neurobehavioral Alterations in Rats Exposed to Silver Nanoparticles

Angela B. Javurek[1], Dhananjay Suresh[2], William G. Spollen[3,4], Marcia L. Hart[5], Sarah A. Hansen[6], Mark R. Ellersieck[7], Nathan J. Bivens[8], Scott A. Givan[3,4,9], Anandhi Upendran[10,11], Raghuraman Kannan[11,12] & Cheryl S. Rosenfeld[4,13,14,15]

Due to their antimicrobial properties, silver nanoparticles (AgNPs) are being used in non-edible and edible consumer products. It is not clear though if exposure to these chemicals can exert toxic effects on the host and gut microbiome. Conflicting studies have been reported on whether AgNPs result in gut dysbiosis and other changes within the host. We sought to examine whether exposure of Sprague-Dawley male rats for two weeks to different shapes of AgNPs, cube (AgNC) and sphere (AgNS) affects gut microbiota, select behaviors, and induces histopathological changes in the gastrointestinal system and brain. In the elevated plus maze (EPM), AgNS-exposed rats showed greater number of entries into closed arms and center compared to controls and those exposed to AgNC. AgNS and AgNC treated groups had select reductions in gut microbiota relative to controls. *Clostridium spp.*, *Bacteroides uniformis*, Christensenellaceae, and *Coprococcus eutactus* were decreased in AgNC exposed group, whereas, *Oscillospira spp.*, *Dehalobacterium spp.*, Peptococcaeceae, *Corynebacterium spp.*, *Aggregatibacter pneumotropica* were reduced in AgNS exposed group. Bacterial reductions correlated with select behavioral changes measured in the EPM. No significant histopathological changes were evident in the gastrointestinal system or brain. Findings suggest short-term exposure to AgNS or AgNC can lead to behavioral and gut microbiome changes.

Engineered silver nanoparticles (AgNPs), which have antimicrobial properties, are increasingly being used in a wide range of consumer products, such as topical wound dressings, coatings for biomedical devices, food-related products to extend the shelf life and reduce potential growth of pathogenic bacteria, and dietary health supplements. Thus, there are mounting concerns of environmental contamination and adverse health consequences[1]. Nanoparticles, owing to their large surface-to-volume ratio, show high reactivity towards cells and easily attach to cellular surfaces. Upon attaching to the surface, nanoparticles can undergo endocytosis, where they may fragment into smaller size nanoparticles with different ionic charges or form aggregates, consequently resulting in compositions with different cellular interaction properties. The size, shape, and surface coating of the nanoparticle determine its cellular adhesion and interaction properties and thereby dictating their toxicity. Besides affecting the host, ingested AgNPs might also impact the gut microbiome.

[1]Department of Occupational and Environmental Health Sciences, West Virginia University, Morgantown, WV, 26506, USA. [2]Department of Bioengineering, University of Missouri, Columbia, MO, 65211, USA. [3]Department of Informatics Research Core Facility, University of Missouri, Columbia, MO, 65211, USA. [4]Bond Life Sciences Center, University of Missouri, Columbia, MO, 65211, USA. [5]Department of Veterinary Pathobiology, University of Missouri, Columbia, MO, 65211, USA. [6]Office of Animal Resources, University of Missouri, Columbia, MO, 65211, USA. [7]Department of Agriculture Experimental Station-Statistics, University of Missouri, Columbia, MO, 65211, USA. [8]DNA Core Facility, University of Missouri, Columbia, MO, 65211, USA. [9]Department of Molecular Microbiology and Immunology, University of Missouri, Columbia, MO, 65211, USA. [10]Department of MU-institute of Clinical and Translational Sciences (MU-iCATS), University of Missouri, Columbia, MO, 65211, USA. [11]Department of Medical Pharmacology and Physiology, University of Missouri, Columbia, MO, 65211, USA. [12]Department of Radiology, University of Missouri, Columbia, MO, 65211, USA. [13]Department of Biomedical Sciences, University of Missouri, Columbia, MO, 65211, USA. [14]Genetics Area Program, University of Missouri, Columbia, MO, 65211, USA. [15]Thompson Center for Autism and Neurobehavioral Disorders, University of Missouri, Columbia, MO, 65211, USA. Angela B. Javurek and Dhananjay Suresh contributed equally to this work. Correspondence and requests for materials should be addressed to A.U. (email: upendrana@health.missouri.edu) or R.K. (email: kannanr@health.missouri.edu) or C.S.R. (email: rosenfeldc@missouri.edu)

A handful of rodent studies have yielded mixed results as to whether AgNPs and other nanoparticles can affect the gut microbiome[2–8]. One study suggests that rats orally exposed for several week duration to various sizes and doses of AgNPs show decreased populations of Firmicutes and *Lactobacillus* but greater proportion of gram negative bacteria, which tend to be more pathogenic in the gastrointestinal system[2]. One limitation of this study is that it only screened select bacterial groups with real-time PCR analysis. Other experiments that employed a global approach by using 16S rRNA sequencing of gut bacteria from mice exposed to similar sizes, doses, and duration of AgNPs reported no effect on gut microbiome communities or diversity[3]. Another study that measured cecal bacterial phyla found that four-week-old rats exposed to varying doses of AgNPs for 28 days did not show any differences relative to controls[8]. A more recent study using next generation sequencing (NGS) demonstrated that mice orally exposed for 28 days to varying doses of AgNPs showed dose-dependent disruptions in types of bacterial sequences (as measured by $\alpha$-diversity) and populations (as measured by $\beta$-diversity)[6]. In this study, exposure to AgNPs increased the ratio between Firmicutes (F) and Bacteroidetes (B) phyla, which was primarily due to changes in Lachnospiraceae and the S24-7 family, respectively.

Effects of AgNPs on the gut microbiome have been examined in other species. In pigs, AgNPs may be a potential antimicrobial additive[9]. Exposure of fruit fly (*Drosophila melanogaster*) larvae to AgNPs results in a less diverse gut microbiota, overgrowth of *Lactobacillus brevis* but a decrease in Acetobacter compared to control or those exposed to copper NPs[10]. Examination of the effects of AgNPs on a defined bacterial community established from a healthy human donor reveals that these particles lead to a negative influence on bacterial communities, as determined by gas production and changes in fatty acid methyl ester profiles[11]. This study also showed AgNPs induce a shift in bacterial communities, as determined by several methods.

*In vitro* and *in vivo* studies also suggest that AgNPs can induce transcriptomic and non-coding RNA changes in neuronal cells and various brain regions, especially the hippocampus[12–20]. Behavioral disruptions, especially in cognitive functions, potential anxiogenic effects, depression-like behaviors, and altered activity levels, have been reported after direct or developmental exposure of rodents and zebrafish to AgNPs[12, 13, 21–25]. Other observed neurotoxicological effects include increased permeability of the blood-brain-barrier (BBB), edema formation, neuronal degradation and apoptosis, synaptic degeneration, tight junction disruptions, increase in reactive oxygen species (ROS), amyloid-$\beta$ (A$\beta$) plagues, and astrocyte swelling[15–18, 20, 22, 23, 26–31].

Potential neurobehavioral disruptions following exposure to AgNPs may be mediated by gut microbiome alterations, i.e. through the gut-microbiome-brain axis. By disrupting the gut microbiome, AgNPs may affect the intestinal epithelial barrier that may increase the likelihood that bacterial pathogens, their virulence factors, and metabolites penetrate to enter the systemic circulation; whereupon, they may reach the brain and cause CNS dysfunction[32–34]. Additionally, AgNPs might be transmitted across the blood-brain-barrier and directly affect various brain regions.

To further understand these potential interactions, we sought to examine whether short-term exposure (two weeks) of Sprague-Dawley rats to different shapes of AgNPs, cube (AgNC) and sphere (AgNS) affected the gut microbiota. Additionally, our aims were to determine whether exposure to AgNC and/or AgNS increased anxiety like behaviors and induced histopathological changes, including increased apoptosis in the amygdala, the primary brain region in rodents and humans governing anxiety-like or stress behaviors[35–39]. The hypotheses of the current studies were that 1) even short-term exposure to one or both forms of AgNPs might result in gut dysbiosis. 2) Oral exposure to AgNPs may directly result in neurobehavioral disturbances, which might also be associated with AgNP-induced gut microbiota composition changes.

## Results

**Characterization of Nanoparticles.** Silver nanoparticles synthesized using polyol process were characterized and tested for stability using transmission electron microscopy (TEM), dynamic light scattering (DLS) and UV-visible spectroscopy. TEM images confirm (Fig. 1) the formation of nanospheres (AgNS) and nanocubes (AgNC). AgNC had an average diameter of 50 nm, while the nanocubes had an edge length of 45 nm. Both particles were highly dispersed and uniform. The surface charge (zeta potential) of the both silver nanocubes and spheres were highly negative, reflecting stable particulate nature. AgNS and AgNC show two sharp characteristic surface plasmon resonance peaks confirming the presence of uniform and stable particles. The physicochemical characterization details are listed in Table 1.

**Stability of Nanoparticles.** The gut and intestinal fluid pH is acidic ranging from 3–7 depending upon the fed or fasted nature[40]. Therefore, prior to *in vivo* testing we performed detailed stability studies in varying pH media and the results are presented below. Nanoparticles were tested for stability in three pH regimes (3, 7 and 9). AgNS and AgNC were monitored for their change in hydrodynamic size, surface charge (zeta potential) and UV-absorption maxima at 24 h and 72 h. The average size of AgNS (93 nm) and AgNC (84 nm) was found to be retained in neutral and basic pH (7 and 9), however a two-fold size increase was observed in pH 3 (Fig. 2a,b). The average surface charge of nanoparticles ($-30$ mV) remained unaltered in basic media, but decreased drastically in both AgNS ($-5$ mV) and AgNC ($-11$ mV) in pH 3 (Fig. 3a,b). AgNS and AgNC showed two characteristic UV-Visible peaks at 355 nm and 420 nm for spheres, and 350 nm and 410 nm for cubes (Fig. 4a,b). The UV-Visible spectra for both AgNS and AgNC showed minimal to no changes in basic pH media but were broad in pH 3. AgNC peaks were relatively narrow compared to AgNS (Fig. 4c,d). The results suggest that although larger aggregates were formed in lower pH media for both AgNS and AgNC, the particulate nature was still retained. The studies further indicate that AgNC has relatively higher stability compared to AgNS, as the changes in size, charge and absorption was comparatively lower.

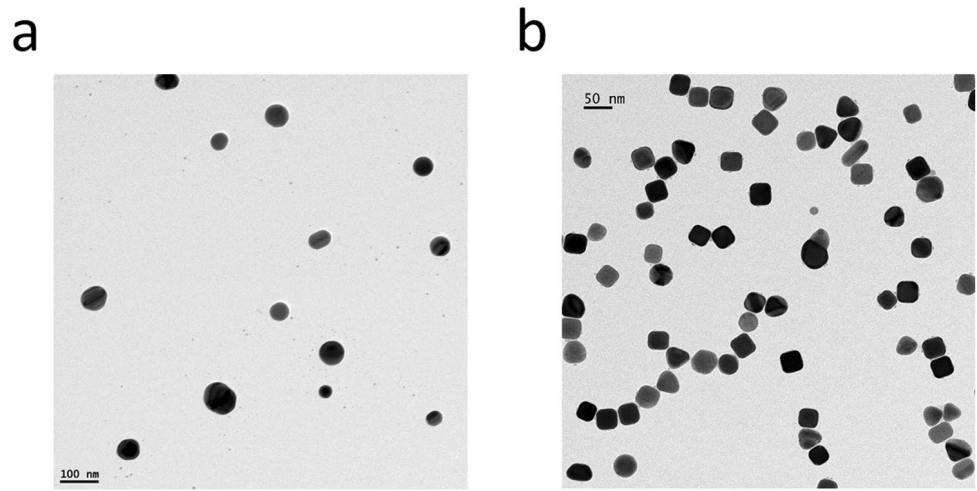

**Figure 1.** TEM images silver nanoparticles (**a**) AgNS and (**b**) AgNC.

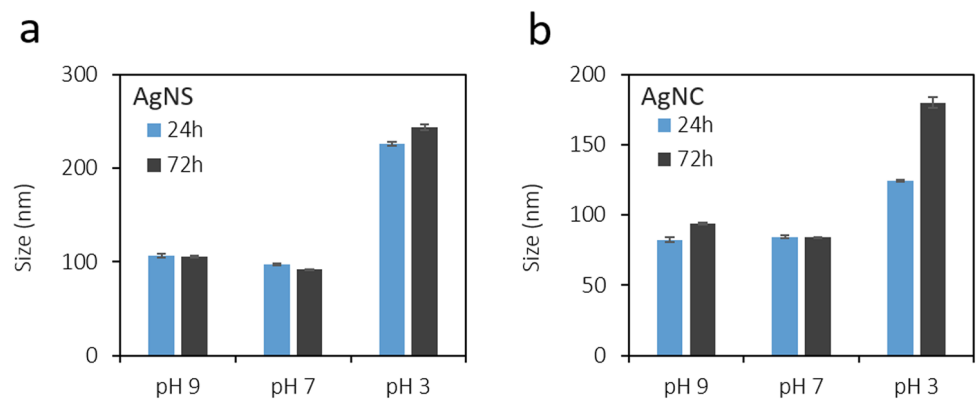

**Figure 2.** Hydrodynamic diameter changes for nanoparticles in different pH conditions: (**a**) AgNS and (**b**) AgNC.

| Silver nanoparticles | Core size | Hydrodynamic size | Surface charge | UV-vis maxima |
|---|---|---|---|---|
| AgNS | 50 nm diameter | 96 nm | −31 mV | 355, 420 nm |
| AgNC | 45 nm edge length | 84 nm | −31 mV | 350, 410 nm |

**Table 1.** Physicochemical characterization of silver nanoparticles.

**Dissolution Studies.** The rats were administered 5 ml of silver nanoconstructs (AgNS and AgNC) at a concentration of 0.2 mg/ml. Therefore, we used the same concentration to perform dissolution experiments as reported in the literature[41]. Dialysis of silver nanoconstructs was performed and the dialysates were used to determine the leaching of silver ions. The amount of silver present in dialysate as quantified by ICP-OES was 0.018 *ppm*.

**Elevated Plus Maze.** Rats exposed to the AgNS had more entries into the closed arms compared to controls and those exposed to AgNC (P = 0.03, Fig. 5A). The AgNS also entered the center more than controls and AgNC-exposed individuals (P values = 0.0002 and 0.03, respectively, Fig. 5B). While there were differences in number of entries into the closed arms and center, there was no difference between any of the groups for the time spent mobile or immobile (Supplementary Figure 1) There was a trend for the AgNS-treated individuals to enter the open arms more than those exposed to AgNC (P = 0.06, Fig. 5C). Both AgNS and AgNC treated rats had increased number of head dipping incidences relative to controls (P = 0.006 and 0.01, respectively, Fig. 5D). None of the other parameters, including duration of time spent in the different arms or center, differed between any of the groups (Data not shown).

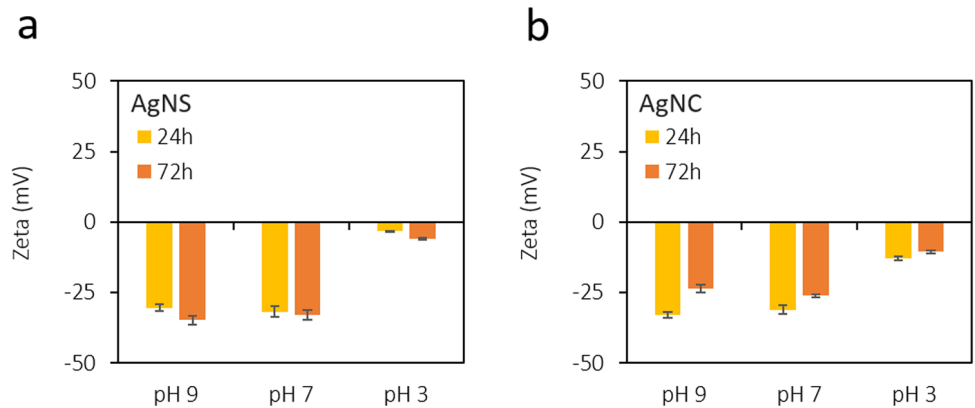

**Figure 3.** Changes in surface charge (zeta potential) for nanoparticles in different pH conditions: (**a**) AgNS and (**b**) AgNC.

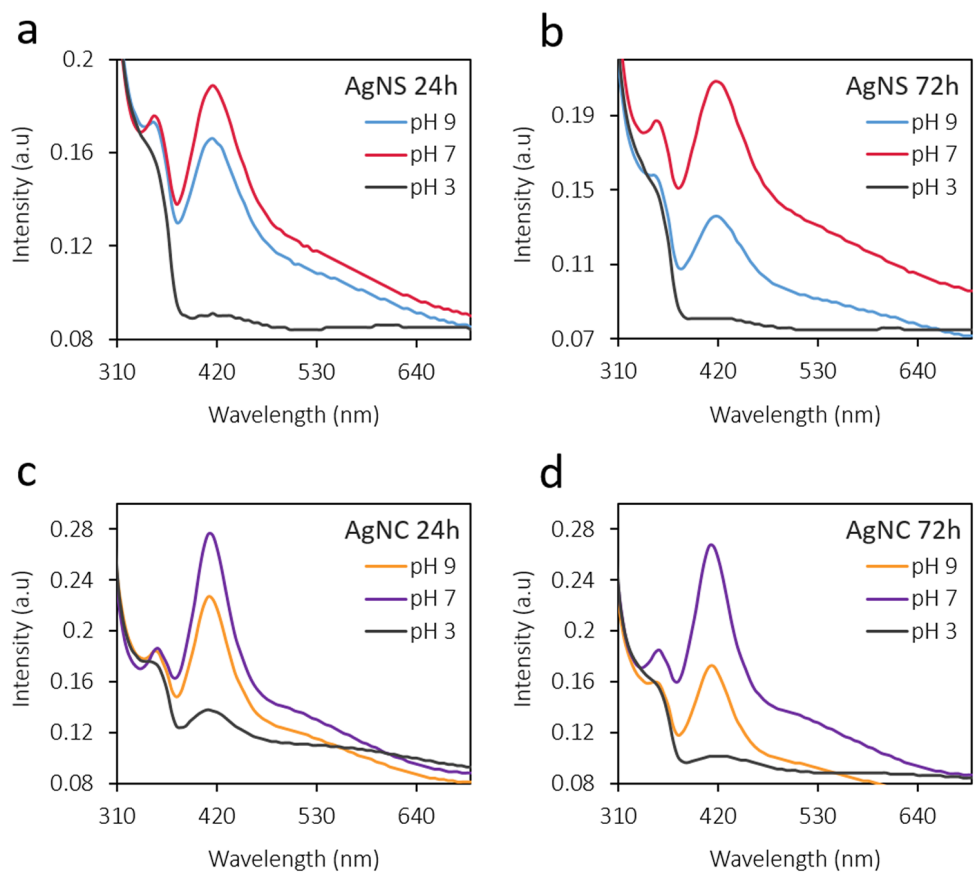

**Figure 4.** UV-visible spectral changes for nanoparticles in different pH conditions: (**a**,**b**) AgNS and (**c**,**d**) AgNC.

**Fecal Microbiome.** Measures of alpha-diversity, including Chao1 and Shannon indices as well as rarefaction analysis, were similar for all of these comparisons (Supplementary Figures 2 and 3). When the 16S rRNA sequencing results were compared using Greengenes Version 13_8 (which is available through QIIME, http://qiime.org/home_static/dataFiles.html ftp://greengenes.microbio.me/greengenes_release/gg_13_5/gg_13_8_otus.tar.gz), no clear distinctions were evident in the various bacterial classes between the three treatment groups (Fig. 6). The PCoA analysis revealed no overt differences between any of the treatment groups (Fig. 7). The PERMANOVA values for the three comparisons were: AgNC vs AgNS P = 0.18; AgNC vs Cont P = 0.50; P = AgNS vs Cont P = 0.26. The identification of the bacterial species included in Fig. 2 is listed in Supplementary Figure 4.

To examine for subtle genera differences, the microbiota data for the three groups were analyzed in a pairwise comparison by using linear discriminant analysis effect size (LEfSe) analysis[42]. *Clostridium spp*., *Bacteroides uniformis*, Christensenellaceae, and *Coprococcus eutactus* were greater in control rats than those treated with AgNC (Fig. 8A). No bacteria were greater in AgNC-exposed rats compared to controls. Comparison of AgNS-treated rats to controls revealed that *Oscillospira spp.*, *Dehalobacterium spp.*, Peptococcaeceae, *Corynebacterium spp.*, *Aggregatibacter pneumotropica* were greater in the latter group (Fig. 8B). No bacteria were greater in the AgNS group relative to controls. Comparison of AgNS to AgNC groups showed that *Anaerostipes spp.*, Rikenellaceae, and Dehalobacteriaeceae were greater in the former group (Fig. 8C). No bacteria were greater in the AgNC group relative to AgNS group.

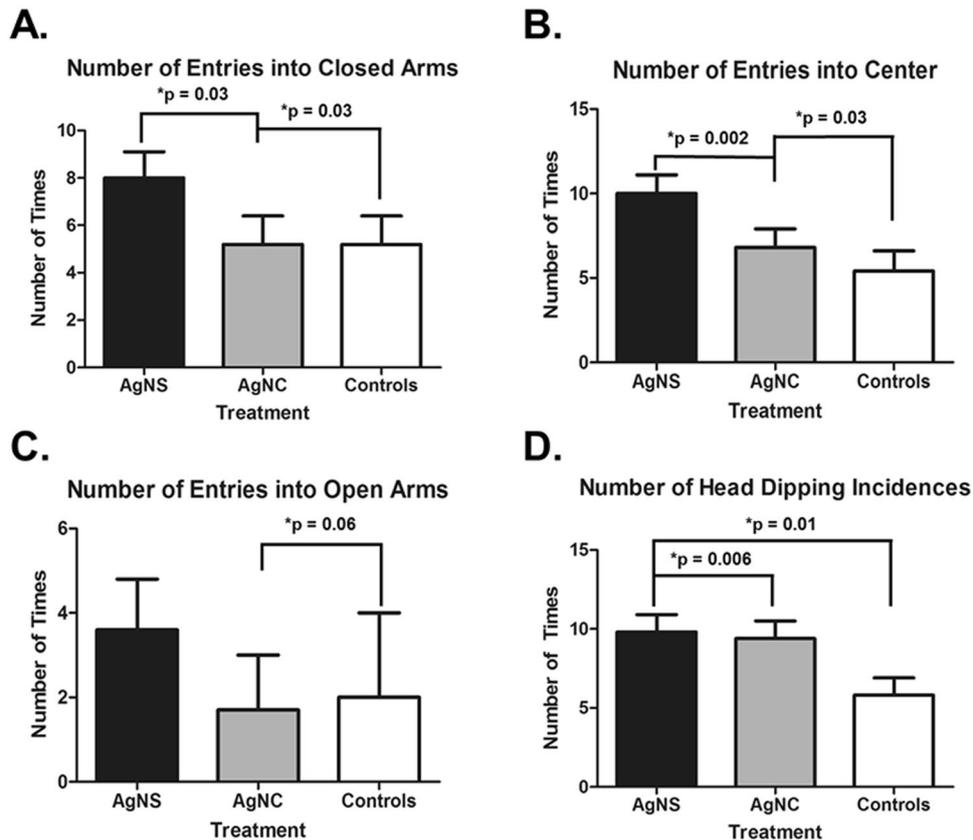

**Figure 5.** Elevated Plus Maze results. (**A**) Number of entries into the closed arms. (**B**) Number of entries into the center of the maze. (**C**) Number of entries into the open arms. (**D**) Number of head-dipping incidences. Significant results are indicated with bars and P values across the treatment comparisons.

Based on the genera that differed in the AgNC/AgNS groups relative to controls, correlation analyses were performed for various KEGG pathways (Figs 9 and 10). Red boxes indicate that the pathways are positively associated with bacterial alterations, whereas blue boxes indicate the pathways are negatively correlated with changes in the various bacteria ($p \leq 0.05$). For instance, decreases in relative abundance of *Bacteroides uniformis* and *Clostridium spp.* in the AgNC group are positively associated with several cancer pathways, biosynthesis of unsaturated fatty acids, arachidonic acid, and α-linolenic acid metabolism (Fig. 9). A decrease in relative abundance of Peptococcaceae in the AgNS group has several positive and negative pathway associations. Examples of positive associations include flavone and flavonol biosynthesis, starch, sucrose and fructose metabolism, and pentose phosphate pathway (Fig. 10). In contrast, many of the amino acid metabolism pathways, streptomycin, penicillin and cephalosporin biosynthesis, β-lactam resistance, citrate cycle (TCA cycle), oxidative phosphorylation, and peptidases are negatively associated with a decrease in this bacterium.

**Histopathology.** To determine if observed behavioral and gut microbiome differences were related to changes at the tissue level, we examined the liver, stomach, small intestine, cecum, and colon for evidence of inflammation, necrosis, and epithelial change (GI tract). No histological differences were noted between control and treatment groups (Fig. 11). In addition to the gastrointestinal tract, we examined the amygdalar region of the brain for evidence of changes in neuronal density (Fig. 12). We found no difference in the number of neurons between treatment and control groups.

**Correlations with Fecal Microbiome-EPM Results.** Correlation of the fecal bacterial changes in the AgNC vs. control groups to the EPM results reveal that a decreased in relative abundance of *Bacteroides uniformis* and *Coprococcus eutactus* in AgNC-treated individuals showed a trend for a negative correlation with time spent in the center of the maze ($P \leq 0.1$, Fig. 13). A decrease in relative abundance of *Clostridium spp.* in this group positively correlated with duration of time spent in the open arms, whereas, this bacterium in this group was negatively associated with number of entries into the closed arms. Changes in the relative abundance of these three bacteria in the controls also correlated with several behavioral results in the EPM, especially for *Bacteroides uniformis* and *Clostridium spp*. Changes in relative abundance of bacteria between AgNS and controls resulted in association trends with the EPM results in the AgNS group (Fig. 14). These include the average time spent in the open arms was negatively associated with a decrease in *Oscillspira spp*. and *Dehalobacterium spp*. (P ≤ 0.1). A decrease in relative abundance of Peptococcaeceae showed a negative association trend with number of times mobile in the EPM, average time spent head dipping, and average time spent rearing, whereas, the number of times rearing showed a trend for a positive association (P ≤ 0.1). In the controls, an increase in *Corynebacterium spp*. was positively associated with duration of time spent immobile (P ≤ 0.05).

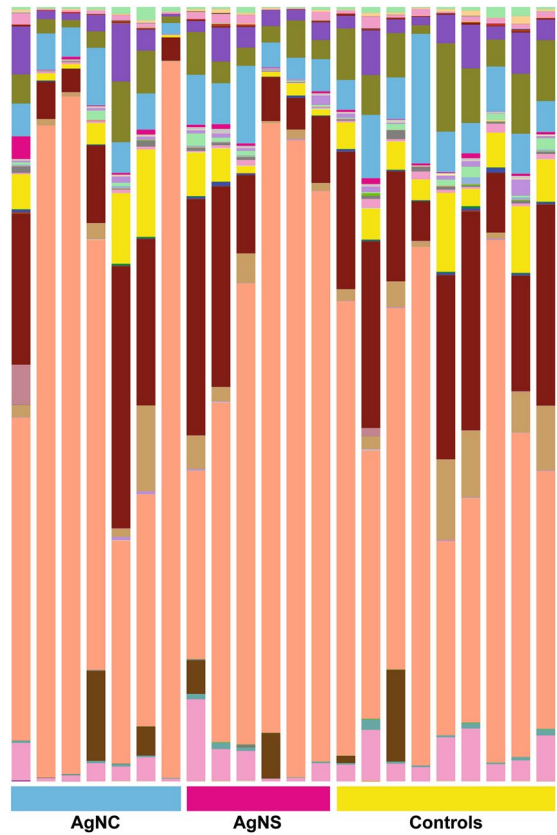

**Figure 6.** Bar plot of fecal microbiome data from AgNC, AgNS, and control groups. Bar plot analysis of the most abundant bacterial classes in all three-treatment groups.

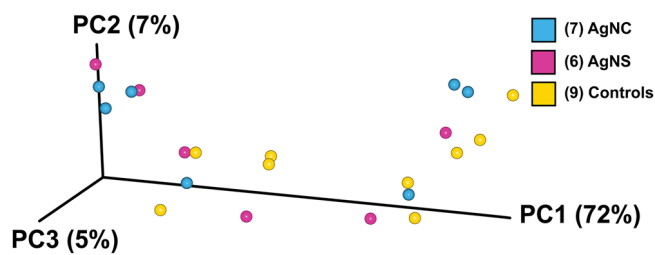

**Figure 7.** PCoA analysis of fecal microbiome data from AgNC, AgNS, and control groups. No clear distinctions are evident based on PCoA analysis. The PERMANOVA values for the three comparisons were: AgNC vs AgNS P = 0.18; AgNC vs Cont P = 0.50; P = AgNS vs Cont P = 0.26.

## Discussion

The goals of the current study were to determine whether short-term exposure (two weeks) of Sprague-Dawley rats to different shapes of AgNPs, cube (AgNC) and sphere (AgNS) increased anxiety-like behaviors. Secondly, we sought to determine whether short-term exposure to these AgNPs resulted in gut dysbiosis. Lastly, we examined whether these AgNPs induced histopathological changes in the gastrointestinal system and brain, including in the amygdala, the primary brain region in rodents and humans governing anxiety-like or stress behaviors[35–39]. Our dissolution studies indicate minimal release of silver ions from the nanoparticles suggesting that the principle action is mediated only by silver nanoparticles.

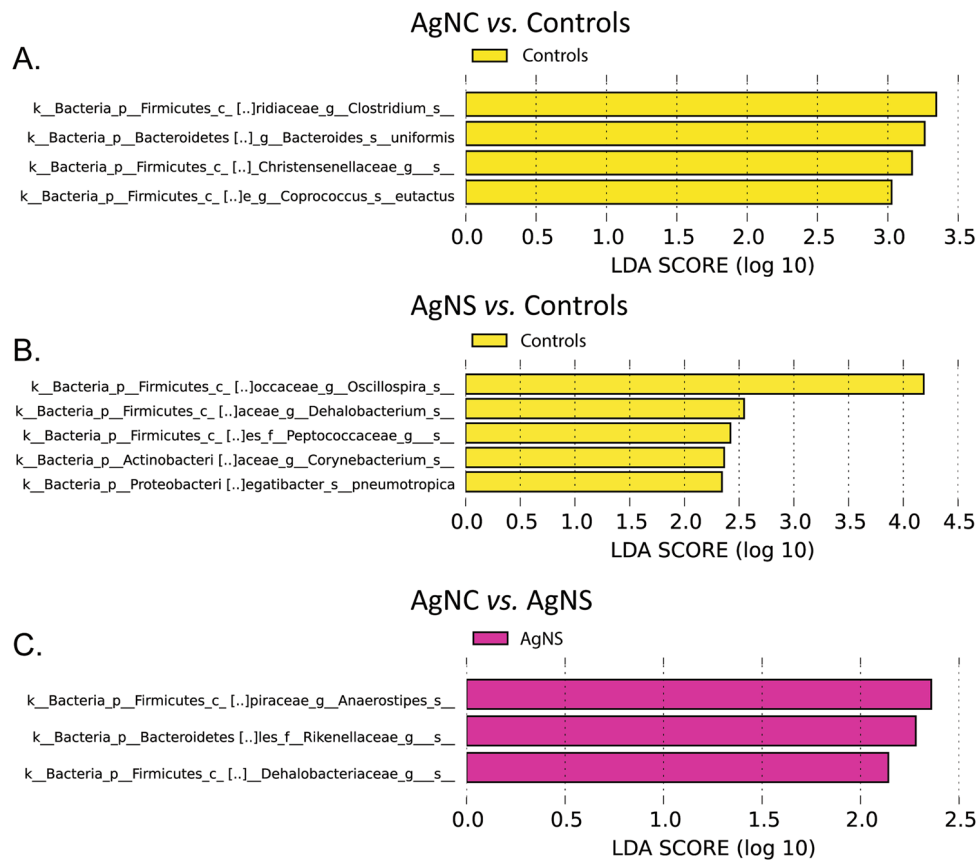

**Figure 8.** LEfSe analysis of fecal microbiome data from AgNC, AgNS, and control groups. (**A**) Comparison of AgNC to controls. The linear discriminant analysis (LDA) score revealed that *Clostridium spp.*, *Bacteroides uniformis*, Christensenellaceae, and *Coprococcus eutactus* were greater in control rats than those treated with AgNC. No bacteria were greater in AgNC-exposed rats compared to controls. (**B**) Comparison of AgNS-treated rats to controls revealed that *Oscillospira spp.*, *Dehalobacterium spp.*, Peptococcaeceae, *Corynebacterium spp.*, *Aggregatibacter pneumotropica* were greater in the latter group. No bacteria were greater in the AgNS group relative to controls. (**C**) Comparison of AgNS to AgNC groups showed that *Anaerostipes spp.*, Rikenellaceae, and Dehalobacteriaeceae were greater in the former group. No bacteria were greater in the AgNC group relative to AgNS group.

For determining the optimal initial dose to test, we used our preliminary studies and previously reported data from the literature[1,13,43,44]. In one study, authors demonstrated that intranasal delivery of both lower dose of 3 mg/kg body weight and higher dose of 30 mg/kg body weight resulted in an increase in reactive oxygen species in the hippocampus and also compromised spatial cognition. On the other hand, other studies have shown that oral dosing of 30 mg/kg body weight of silver nanoparticles (AgNP) resulted in toxicity in major organs[2]. However, these other reports are largely limited to understanding whether these particles result in whole organ damage. As our current study was designed to understand the effects of AgNPs on the gut microbiome and relation to behavioral outcomes, we chose the lower dosing of 3.6 mg/kg, which also better aligns with the aforementioned spatial cognition study.

The current studies suggest that rats exposed to AgNS might be somewhat more anxious, as evidenced by increased entries into closed arms, compared to controls and those exposed to AgNC. However, this group also entered the center more than controls. Both AgNS and AgNC treated rats engaged in more head-dipping incidences, which might be considered a repetitive behavior. Head-dipping behavior was measured as potential repetitive behavior since it is a behavior that rodents engage in commonly while in the EPM[45–48]. This behavior may also be an indicator of anxiety-like state with some studies suggesting an anxiolytic[45,46] but others suggesting anxiogenetic state[47,48]. Thus, it is not fully clear at this point whether this behavioral change represents an increase in stereotypical behaviors or an indicator of anxiety-like state in these treated groups. Additional studies with varying dose ranges are needed to confirm the significance of the behavioral findings identified in the EPM. A study with NMRI mice suggests that prenatal exposure to AgNP may not affect anxiety-like behaviors when tested in the EPM; however, number of defecations in the open field assay and number of passages in the light-dark box are increased in prenatally exposed mice[21]. Other behavioral domains in rats, especially, spatial learning and memory, as assessed in the Morris Water Maze (MWM), appear to be compromised after exposure to AgNPs[12,13].

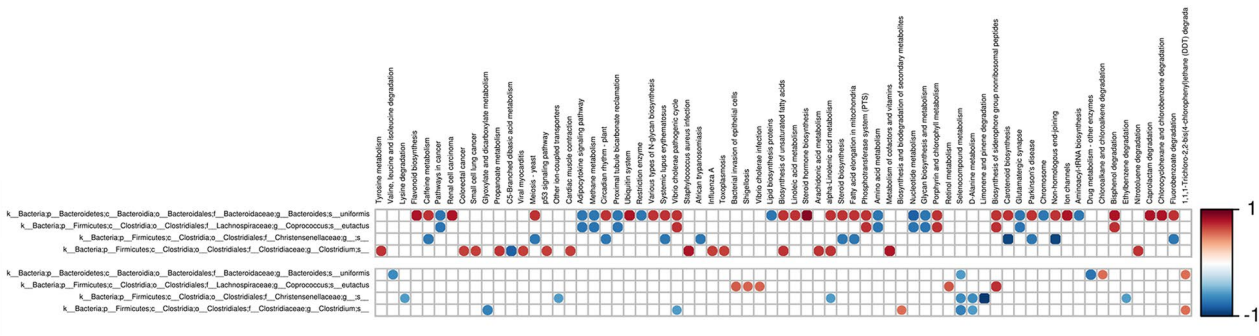

**Figure 9.** Bacterial metabolic and other pathway differences in the fecal samples of AgNC-treated individuals vs. controls. As described in Fig. 7 of previously published article[80] and in the Materials and Methods section, correlations between the PICRUSt-generated functional profile and QIIME-generated genus level bacterial abundance were calculated and plotted against treatment. Those genera with a LEfSe LDA score ≥2 between controls and BPA individuals are depicted. Metabolic pathway designations are delineated at the bottom of the figure. Shading intensity and size of the circles indicates the Kendall rank correlation coefficient between matrices. Orange/red indicates a positive correlation; whereas blue designates a negative correlation at a p value ≤ 0.05.

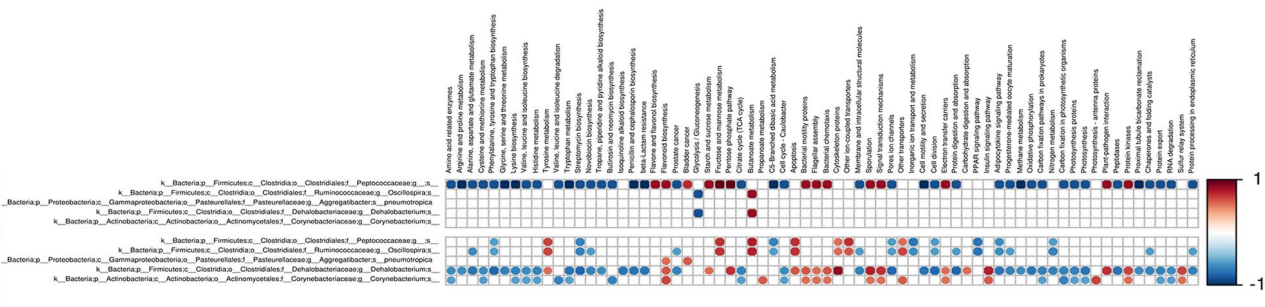

**Figure 10.** Bacterial metabolic and other pathway differences in the fecal samples of AgNS-treated individuals vs. controls. Data are presented as detailed in Fig. 9.

Depression-like behaviors, and altered activity levels, have been also reported after direct or developmental exposure of rodents and zebrafish to AgNPs[22–25].

In the gut microbiome assessments, no bacteria were elevated in the AgNS or AgNC groups relative to controls. However, the relative abundances of select bacteria were reduced in rats treated with either AgNPs compared to the control group. In the AgNC group, *Clostridium spp.*, *Bacteroides uniformis*, Christensenellaceae, and *Coprococcus eutactus* were decreased, whereas, *Oscillospira spp.*, *Dehalobacterium spp.*, Peptococcaeceae, *Corynebacterium spp.*, *Aggregatibacter pneumotropica* were reduced in AgNS group. Other studies have yielded conflicting findings as to whether exposure to AgNPs or other nanoparticles can affect the gut microbiome[2–7]. In rats, several weeks duration to various sizes and doses of AgNPs decrease populations of Firmicutes and *Lactobacillus* but results in elevations of gram negative and likely pathogenic bacteria[2]. One limitation is that this study used real-time PCR analysis to screen and select bacterial groups. However, 16S rRNA sequencing analyses of the gut microbome in mice exposed to AgNPs vs. controls did not show any differences[3]. Another study that measured cecal bacterial phyla found that four week old rats exposed to varying doses of AgNPs for 28 days did not show any differences relative to controls[8]. By using next generation sequencing (NGS), another group demonstrated that mice orally exposed for 28 days to varying doses of AgNPs exhibit dose-dependent disruptions in both α and β-diversity of the gut microbiome[6]. Further, the ratio between Firmicutes (F) and Bacteroidetes (B) phyla was greater in the exposed group with shift primarily due to changes in Lachnospiraceae and the S24-7 family, respectively. The conflicting results across our current and past studies may be due to rodent model tested, method used to assess the gut microbiome, dose and physical property of the AgNPs, and duration of exposure. AgNPs may also affect other vertebrate and invertebrate species[9–11].

In the current studies, we did not detect any histopathological changes in the gastrointestinal system and brain regions, including number of neurons on the amygdala. It could be that longer-term and higher dose exposure to AgNPs might cause histopathological changes. The possibility that this short-term regimen resulted in ultra-structural changes cannot be ruled out. Past studies with varying length to AgNP exposure in rodents and fish suggest that these chemicals can result in several neuropathological changes, such as edema formation, neuronal degradation and apoptosis, synaptic degeneration, tight junction disruptions, increase in reactive oxygen species (ROS), disturb brain antioxidant system, amyloid-β (Aβ) plagues, and astrocyte swelling[15–18, 20, 22, 23, 26–31, 49–52].

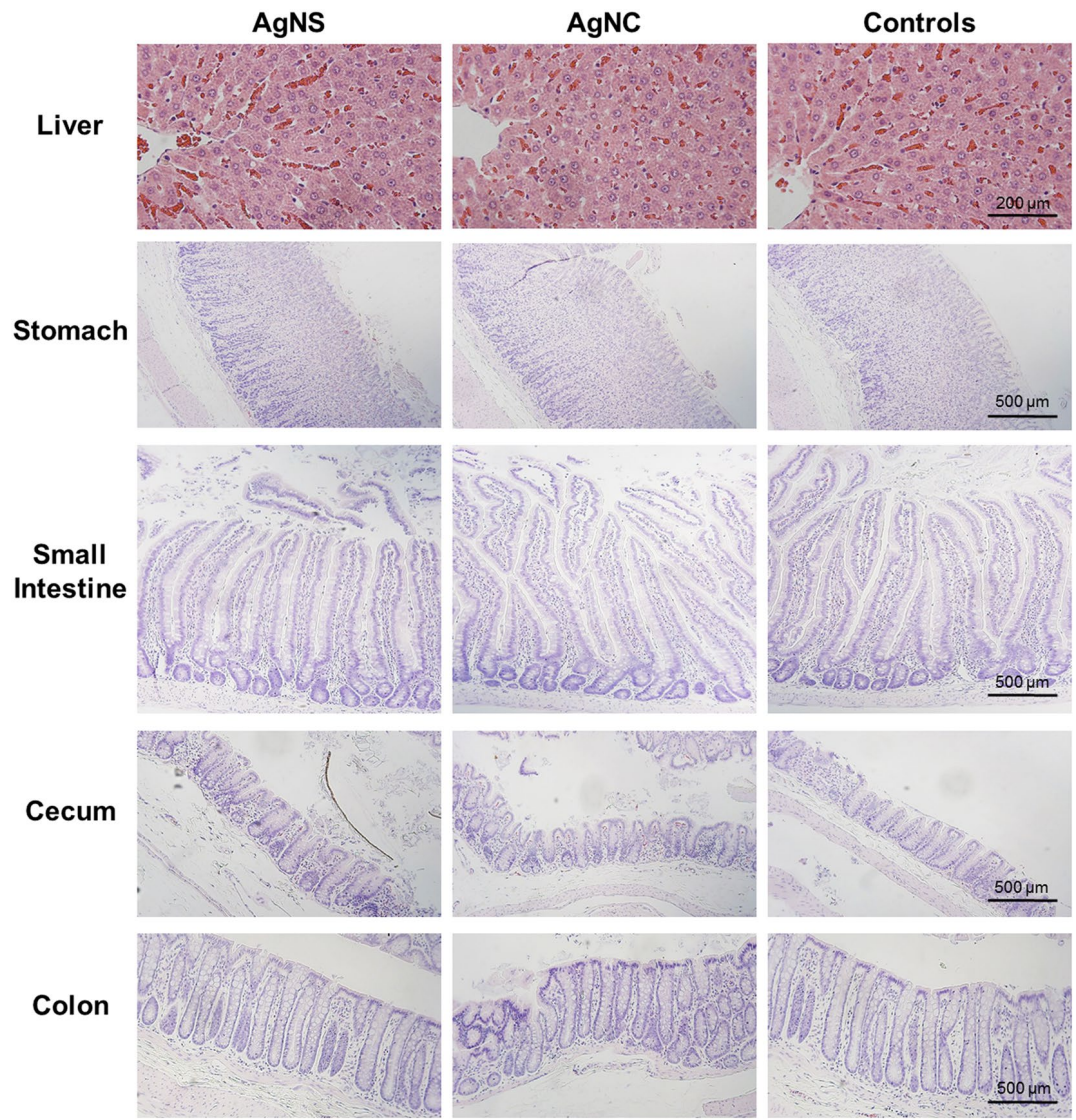

**Figure 11.** Comparison of liver and gastrointestinal tract following administration of silver nanoparticles (AgNC and AgNS). H&E stained representative tissue sections. Bar = 200 μm for liver and 500 μm for GI tract.

AgNPs may also disrupt the cytoskeleton architecture in neuronal cells[53]. Several transcripts are altered in neurons of mice or rats exposed to AgNPs, such as *Psen1*, *Psen2*, *Trex1*, *Irf7*, *RasGrf1*, *Bcl2*, *Th*, *Maoa*, *Cdh1*, *Cldn1*, and *Il4*[16, 54, 55].

The observed behavioral changes in the EPM might be due to the possible molecular and ultrastructural changes mentioned above. Another possibility is that the behavioral changes are in part due to AgNP disruption in the gut microbiome. Substantial evidence supports the existence of a gut-microbiome-brain axis[34, 56–58]. The original findings supporting such an axis come from gnotobiotic or germ-free (GF) mice that lack a gut microbiome and demonstrate several behavioral changes relative to counterparts possessing this microbiome, especially increased anxiety-like behaviors[59–64]. Other animal model and human epidemiological data that indicate shifts in the gut microbiota populations are associated with neurobehavioral disorders in rodent models and humans, in particular autism spectrum disorders[65–70].

In the current results, relative reductions in select bacteria in the AgNC or AgNS either positively or negatively correlated with behaviors measured in the EPM. These findings do not indicate that shifts in these isolated bacteria alone are responsible for mediating these behaviors, only that changes in these bacteria are associated and may partially contribute to the observed behavioral changes. In other words, we cannot establish causation based on these studies alone.

In addition to the above, the body and organ weights of the treated and control groups were also compared. Analysis of the weights show no significant differences (Supplementary Figure 5), suggesting that there was minimal to no overt toxicity caused in these animals due to AgNS or AgNC exposure.

Future work will extend the length of exposure and test a range of doses to determine if chronic exposure results in heightened gut microbiome and neurobehavioral alterations, such as possible cognitive and social deficits. Transmission electron microscopy will be used to assess for ultrastructural changes induced by AgNPs and the amount of distribution of these particles to the brain after oral exposure. Moreover, additional analyses will be done on the gastrointestinal system to examine for potential increased gut leakiness and enterocyte alterations. Future studies will analyze the global transcriptomic profile in the intestine and brain regions modulating anxiety-like and social behaviors, learning and memory. Finally, in these initial studies, only males were examined to reduce potential confounding effects of the estrous cycle in females. Follow-up work will examine effects in both males and females where vaginal cytology will be done to ensure that the females are at the same stage of the estrous cycle at the initiation of the studies.

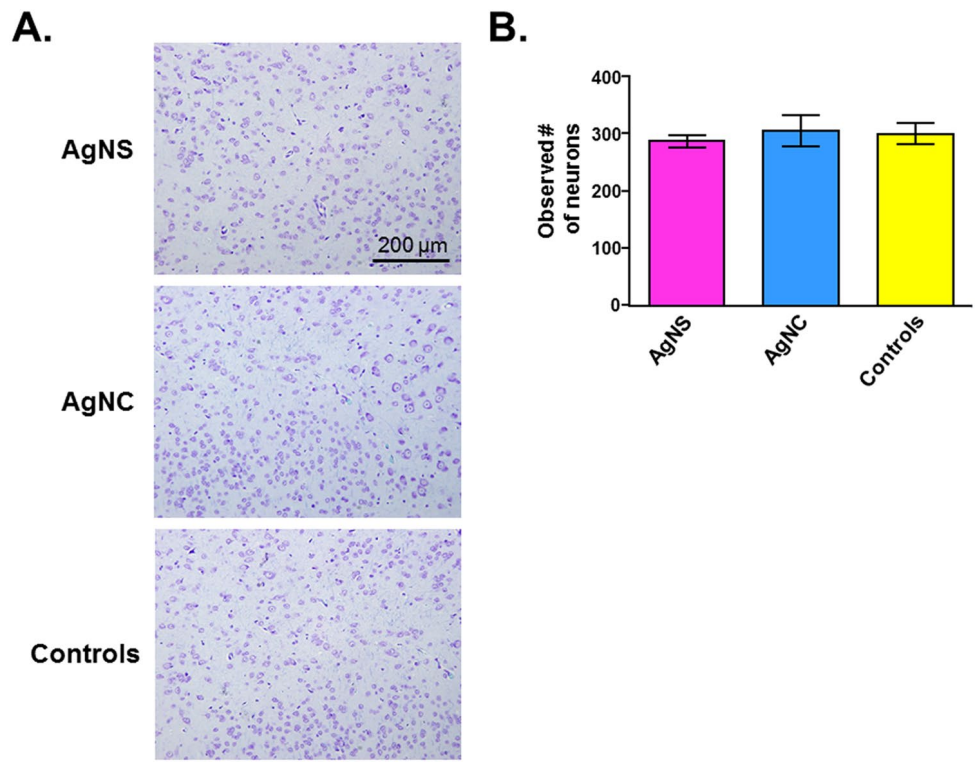

**Figure 12.** Comparison of neuronal density in the amygdala. (**A**) Cresyl violet-luxol fast blue staining of representative tissue sections. (**B**) Number of observed neurons per five 200x fields of view. Values are the mean ± SEM.

In conclusion, the current findings indicate that short-term and low dose exposure of male rats to AgNPs can result in an increase in anxiety-like and possibly stereotypical behaviors with effects being more pronounced in those exposed to AgNS. Both AgNC and AgNS demonstrated antimicrobial activity by reducing select bacteria comprising the gut microbiome. The bacterial changes correlated with some of the behaviors measured in the EPM. No evidence of overt histopathological changes was identified in the gastrointestinal system or brain. However, it remains to be determined whether the identified behavioral changes are due to AgNP-induced transcriptomic and ultrastructural changes in the brain versus the contribution that occurs secondary to gut dysbiosis. With our increasing use of AgNPs, the current studies suggest some cause for concern about potential exposure in humans, although future studies are needed before any firm conclusions can be drawn on their toxicity to the host and gut microbiome.

## Methods

**Synthesis of Nanospheres.** Silver nanospheres (AgNS) coated with polyvinylpyrrolidone (PVP) was prepared according to the described procedures. A round bottom flask containing 60 ml of ethylene glycol was heated at 155°C for 55 min at 400 RPM. After 55 min, argon was bubbled and 0.7 ml of sodium hydrosulfide (3 mM) and 15 ml of PVP (26.66 mg/ml; MW 55 000) dissolved in ethylene glycol was added quickly to the round bottom flask. After 8 min, 5 ml of $AgNO_3$ (48 mg/ml) dissolved in ethylene glycol was quickly added to the reaction solution. After 25 min, the system was introduced to air, and then the reaction was stopped by cooling the solution to room temperature. The color of the final solution was greyish brown with a distinct red meniscus. To the final solution, 40 ml of acetone was added, centrifuged (6000 g for 20 min) to obtain the pellet. The pellet was washed with sterile DI water six times (7000 g for 20 min) to remove unbound PVP. The pellet was resuspended in sterile DI water and stored at room temperature (0.2 mg/ml). The particles were characterized by zetasizer and TEM measurements.

**Synthesis of Silver Nanocubes.** Silver Nanocubes (AgNC) were synthesized with a slight modification of a previously published procedure[71]. One modification included that argon bubbling of sodium hydrosulfide, PVP and AgNO$_3$ solutions was performed for 5 min prior to addition to ethylene glycol solution. After 20 min of AgNO$_3$ addition, green ochre solution was obtained and the reaction was stopped by cooling to room temperature under argon. The reaction mixture was then mixed with 40 ml of acetone, centrifuged (6000 g for 20 min) to obtain a pellet. The pellet was further washed with sterile DI water 6 times (7000 g for 20 min) to remove unbound PVP. Final solution was resuspended in sterile DI water (0.2 mg/ml) and stored at room temperature. The particles were characterized by zetasizer and TEM measurements.

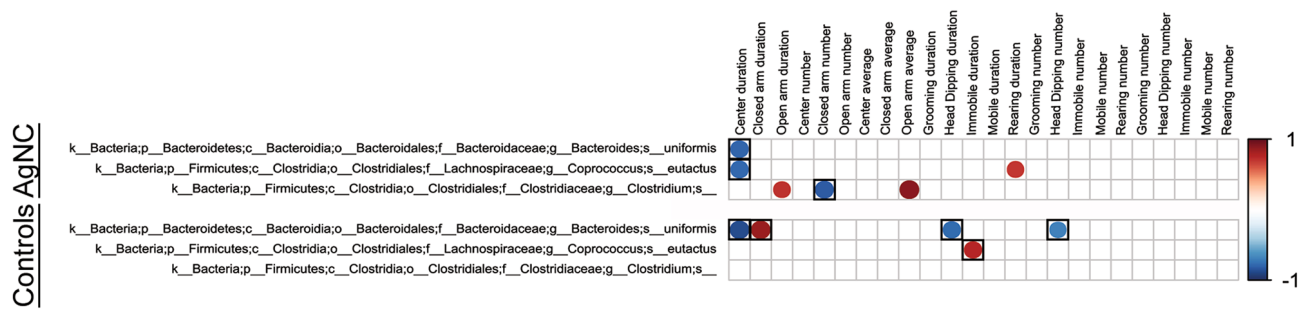

**Figure 13.** Correlations of fecal microbiota changes and EPM results in AgNC and control groups. This figure correlates the bacterial changes identified with LEfSe (Fig. 8) and the EPM results (Fig. 5). The correlations and data presented as detailed in Fig. 9. However, the values without a box only showed a statistical trend (P ≤ 0.1 for significance), whereas those with a black box were significant (P ≤ 0.05).

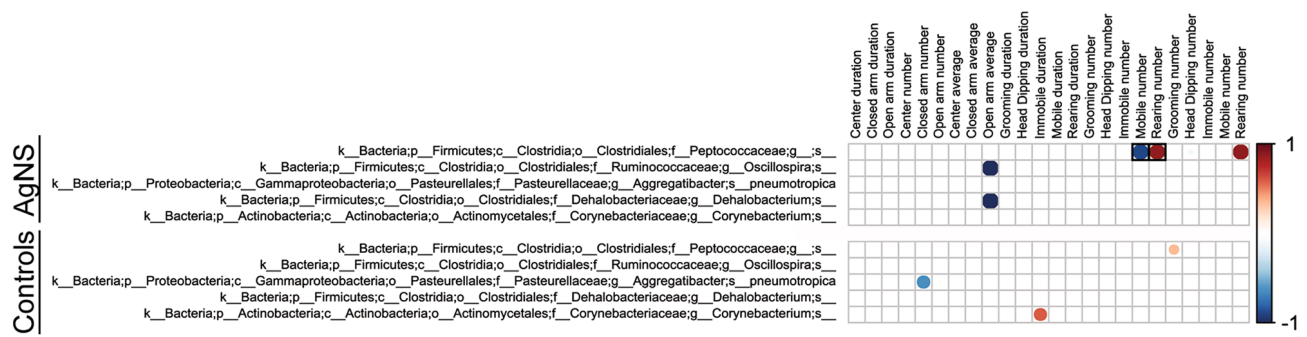

**Figure 14.** Correlations of fecal microbiota changes and EPM results in AgNS and control groups. This figure correlates the bacterial changes identified with LEfSe (Fig. 8) and the EPM results (Fig. 5). The correlations and data presented as detailed in Fig. 9. However, the values without a box only showed a statistical trend (P ≤ 0.1 for significance), whereas those with a black box were significant (P ≤ 0.05).

**Dissolution Experiments.** For this experiment, a DiaEasy Dialyzer Tube was first equilibrated with water. 1 ml of silver nanoconstructs (AgNS and AgNC (0.2 mg/ml)) dissolved in water was added to the dialyser tube. The tube was then immersed in 20 ml of water in a beaker and dialysis was performed for 48 h at 26 °C. The dialysate was concentrated to 2 ml using a rotary evaporator and analyzed for silver ions. Duplicate measurements were performed. Silver standard routinely used for ICP-OES analysis was used for calibration. The amount of silver present in dialysate was quantified by ICP-OES.

**Animals and Exposure to Nanoparticles.** All experiments were approved by University of Missouri Animal Care and Use Committee (Protocol #8639) and performed in accordance with the recommendations in the NIH Guide for the Care and Use of Laboratory Animals[72]. Seven week old Crl:CD(SD) male rats were procured from Charles River Laboratories and assigned to one of three daily oral treatments (n = 9 per treatment group) administered via 18 gauge, 75 mm long disposable gavage needles (Instech, Plymouth Meeting, PA): (1) 3.6 mg/kg cube shaped silver nanoparticles (AgNC) in sterile water; (2) 3.6 mg/kg body weight sphere shaped silver nanoparticles (AgNS) in sterile water; or (3) sterile water. The total volume administered was 20 ml/kg, and the rats were treated for fourteen days. Male rats were used in this initial experiment to minimize potential confounding effects of the estrous cycle. Animals receiving the same treatment were housed together to avoid any cross-contamination. Animals were housed at the University of Missouri in sterile cages in a room maintained at 68–79 °F, average humidity of 30–70% with a 12:12 hr light:dark cycle, and provided *ad libitum* acidified water and rodent chow (LabDiet 5008, St. Louis, MO). Routine sentinel testing confirmed that rats were free of Sendai virus, Pneumonia Virus of Mice (PVM), sialodacryoadentitis virus, Kilham rat virus, Toolan's H-1 virus, rat parvovirus, rat minute virus, parvovirus NS-1, Theiler's-like virus of rats, *Mycoplasma pulmonis*, *Aspiculuris tetraptera*, and *Syphacia muris* (Charles River Laboratories). The University of Missouri is accredited by the Association for Assessment and Accreditation of Laboratory Animal Care (AAALAC International).

**Elevated Plus Maze (EPM) Testing.** After the rats had been exposed to the silver nanoparticles for fourteen days, they were tested in the EPM to measure exploratory and anxiety-like behaviors. This procedure was performed as described previously[73–76]. Briefly, the EPM is arranged in a plus configuration and includes two opposite open arms (30 cm), a central platform region (5 × 5 cm), and two opposite closed arms (30 cm). Each animal was placed in the center of the maze and permitted to explore it for 300 seconds. Each trial was recorded with a Canon Vixia HF HD hand held camcorder (Canon). The video trials were then analyzed with the Observer Version 11 software (Noldus Technologies, Leesburg, VA). Parameters measured include duration of time spent and frequency entering the open and closed arms, center of the maze, duration and number of times engaging in head-dipping and rearing behaviors.

**Fecal Collection.** The day before the rats were tested in the EPM (day 13 of treatment), they were placed in individual cages without any bedding material and four to five fecal boli were collected from each animal. They were only in individual cages for as long as it took to collect this number of fecal boli, which was about 15 to 20 minutes, and thus, this amount of time spent in an individual cage should have had minimal effects on anxiety. Fecal boli were collected immediately after defecation, and the tubes containing them placed on dry ice. After fecal collection, the animals were returned to their home cages with other animals receiving similar treatments. The fecal samples were transported on dry ice and then stored at −80 °C until they were processed for bacterial DNA isolation, detailed below. The amount of time from collection to isolation was approximately two to three weeks. However, it is uncertain whether this amount of time would have dramatically influenced the microbial populations, especially as the samples were stored during this time at −80 °C.

**Fecal Bacterial DNA Isolation.** This procedure was performed as we have previously described[77–79]. Briefly, fecal microbial DNA was isolated from all nine replicates per treatment group using the PowerFecal DNA Isolation kit (Mo Bio Laboratories, Inc., Carlsbad, CA) and in accordance with the manufacture's protocol. Quantity of bacterial DNA was measured using Qubit 3.0 Fluorometer (Life Technologies, Grand Island, NY). The number of replicates tested is comparable to those used in other gut microbiome studies, including our recent study examining the effects of BPA exposure in a rodent model, that have shown such sample sizes can result in statistical differences between groups[77, 80, 81].

**16S rRNA Sequencing of Bacterial DNA from Fecal Samples.** The University of Missouri DNA Core Facility prepared bacterial 16S ribosomal DNA amplicons from extracted fecal DNA by amplification of the V4 hypervariable region of the 16S rDNA as we have previously described[77–79]. Final amplicon pool was evaluated using the Advanced Analytical Fragment Analyzer automated electrophoresis system, quantified with the Qubit flourometer using the quant-iT HS dsDNA reagent kit (Invitrogen), and diluted according to Illumina's standard protocol for sequencing on the MiSeq. Sequence data was generated using a paired-end, 250 base pair read length.

**Bioinformatics and Amplicon Analyses of 16s rRNA Sequencing Results.** Paired-end Illumina MiSeq DNA reads were joined using FLASH[82] and cutadapt (v 1.8.3[83]) was used to remove primers. Usearch7[84] was used to remove those contigs with more than 0.5 expected number of errors (http://drive5.com/usearch/manual/exp_errs.html). The uparse algorithm[85] was used to cluster de novo the contigs to 97% identity, and remove suspected chimera. A representative sequence from each cluster was annotated against DNA sequences in the Greengenes database (v 13_8[86]). For alpha-diversity, Chao1 (species richness) and Shannon (species diversity) values were calculated and plotted using the R package phlyoSeq[87]. Rarefaction metrics were calculated using the alpha_rarefaction.py script in the QIIME package[88]. Measurements of beta-diversity were performed by the QIIME script beta_diversity_through_plots.py.

Closed reference OTU picking to 97% identity was performed using QIIME script pick_closed_reference_otus.py against the Greengenes database (v 13_5). This was done to allow subsequent use of the PICRUSt[89] software to predict the functional content of the metagenome from the 16S rRNA sequences. The resulting taxa were collated into 99 distinct taxonomic groups using the QIIME script summarize_taxa.py. All subsequent analyses were performed using this closed-reference OTU table. LEfSe[42] was used to identify taxa most characteristic of the three treatment groups, and significant taxa were visualized as a bar-chart of the effect relevance. Bacterial metabolic characterization of sample types was facilitated with PICRUSt, version 1.0.0. To correlate the taxa abundance with metabolic characteristics of sample types, a custom R script provided as a gift from Dr. Jun Ma and Kjersti Aagaard-Tillery, Baylor College of Medicine, Houston, TX was used. In these figures, the correlation of the abundance of taxa (from the OTU table) with their predicted metabolic function (from KEGG pathways as determined by PICRUSt), was calculated with the R stats function cor.test (https://cran.r-project.org/), using the Kendall method, a rank-based measure of association. The cor.test function outputs the correlation coefficient and significance of a comparison of an OTU with a KEGG term across samples. The matrix of all the correlation values was visualized using the R package corrplot (https://cran.r-project.org). The area and intensity change together so that larger, darker, circles represent correlation coefficients that are larger in magnitude. The scale to the right of each figure relates those shades of color to the value of the correlation coefficient. Up to 70 of the most abundantly represented KEGG terms that had a significant (alpha < 0.05) correlation with one or more taxa were included in the plot. Correlation of taxa abundance with the EPM was done with Kendall method.

**Histopathological Analyses and Morphometric Quantification.** At the end of the study, rats were individually euthanized with $CO_2$ at a displacement rate of 20% of chamber volume/minute controlled by a flow meter. Cardiocentesis was performed, followed by saline perfusion, then 10% neutral-buffered formalin (NBF) perfusion to preserve the tissues. Liver, stomach, small intestine, cecum, colon, and brain were removed, immersed in 10% buffered formalin for 48 hr and processed for hematoxylin and eosin, and cresyl violet-luxol fast blue staining. Histological evaluation was performed by a rodent pathologist, blinded to the identity of the animal's experimental group.

**Cresyl violet-Luxol fast blue staining of the Amygdala and morphometric quantification.** Brain samples were trimmed at level three using gross anatomical landmarks as described previously[90]. Sections were evaluated for neuronal staining with cresyl violet and density evaluated in five 200x fields of view per animal. Statistical analysis was performed using Sigma Plot 13.0 (Systat Software Inc., Carlsbad CA). Differences in neuronal density were determined using one-way ANOVA followed by Student Newman-Keuls post hoc test. A $p$ value < 0.05 was considered statistically significant.

**Statistics.** SAS version 9.2 software analyses software (SAS Institute, Cary, NC) was employed for these analyses. Unless otherwise stated, the reported data are based on mean ± SEM. For the EPM testing, the amount of total time spent in the open and closed arms and center, as well as total number of arm entries, average velocity, total distance travelled, number of times and duration engaged in head-dipping and rearing were analyzed by ANOVA.

### Acknowledgements
The authors are grateful for assistance from Sarah A. Johnson who helped with the maze design, Wayne Shoemaker who constructed the elevated plus maze, Sherrie Neff and residents in lab animal pathology who assisted with the tissue collections, and Donald E. Connor who helped with the figure preparations.

### Author Contributions
A.B.J., D.S., A.U., R.K., and C.S.R. designed the study, A.B.J., D.S., M.L.H., S.A.H., and N.J.B. performed the experiments, W.G.S., M.L.H., M.R.E., and S.A.G. analyzed the data, and A.B.J., D.S., W.G.S., M.L.H., S.A.H., M.R.E., N.J.B., S.A.G., A.U., R.K., and C.S.R. prepared the manuscript. All authors reviewed the manuscript.

### Additional Information
**Supplementary information** accompanies this paper at doi:10.1038/s41598-017-02880-0

**Competing Interests:** The authors declare that they have no competing interests.

**Publisher's note:** Springer Nature remains neutral with regard to jurisdictional claims in published maps and institutional affiliations.